\documentclass{vgtc}                          % final (conference style)
\ifpdf%                                % if we use pdflatex
  \pdfoutput=1\relax                   % create PDFs from pdfLaTeX
  \usepackage{graphicx}                % allow us to embed graphics files
  \DeclareGraphicsExtensions{.pdf,.png,.jpg,.jpeg} % for pdflatex we expect .pdf, .png, or .jpg files
\else%                                 % else we use pure latex
  \usepackage{graphicx}                % allow us to embed graphics files
  \DeclareGraphicsExtensions{.eps}     % for pure latex we expect eps files
\fi%

\graphicspath{{figures/}{pictures/}{images/}{./}} % where to search for the images

\usepackage{microtype}                 % use micro-typography (slightly more compact, better to read)
\PassOptionsToPackage{warn}{textcomp}  % to address font issues with \textrightarrow
\usepackage{textcomp}                  % use better special symbols
\usepackage{mathptmx}                  % use matching math font
\usepackage{times}                     % we use Times as the main font
\usepackage{cite}                      % needed to automatically sort the references
\usepackage{tabu}                      % only used for the table example
\usepackage{booktabs}                  % only used for the table example
\onlineid{1147}

\vgtccategory{Application}

\title{Vis-SPLIT: Interactive Hierarchical Modeling for mRNA Expression Classification}

\author{%
  \authororcid{Braden Roper}{0000-0002-4802-9473}
  \thanks{e-mail: bradenroper@ou.edu}\\%
  \scriptsize University of Oklahoma %
  \and \authororcid{James C.\ Mathews}{0000-0002-8247-7536}
  \thanks{e-mail: mathewj2@mskcc.org}\\%
  \parbox{1.4in}{\scriptsize \centering Memorial Sloan Kettering \\ Cancer Center}
  \and \authororcid{Saad Nadeem}{0000-0001-5030-9792}
  \thanks{e-mail: nadeems@mskcc.org}\\%
  \parbox{1.4in}{\scriptsize \centering Memorial Sloan Kettering \\ Cancer Center}
  \and \authororcid{Ji Hwan Park}{0000-0002-7971-2419}
  \thanks{e-mail: jpark@ou.edu}\\%
  \scriptsize University of Oklahoma %
}

\teaser{
  \centering
  \includegraphics[width=0.99\linewidth]{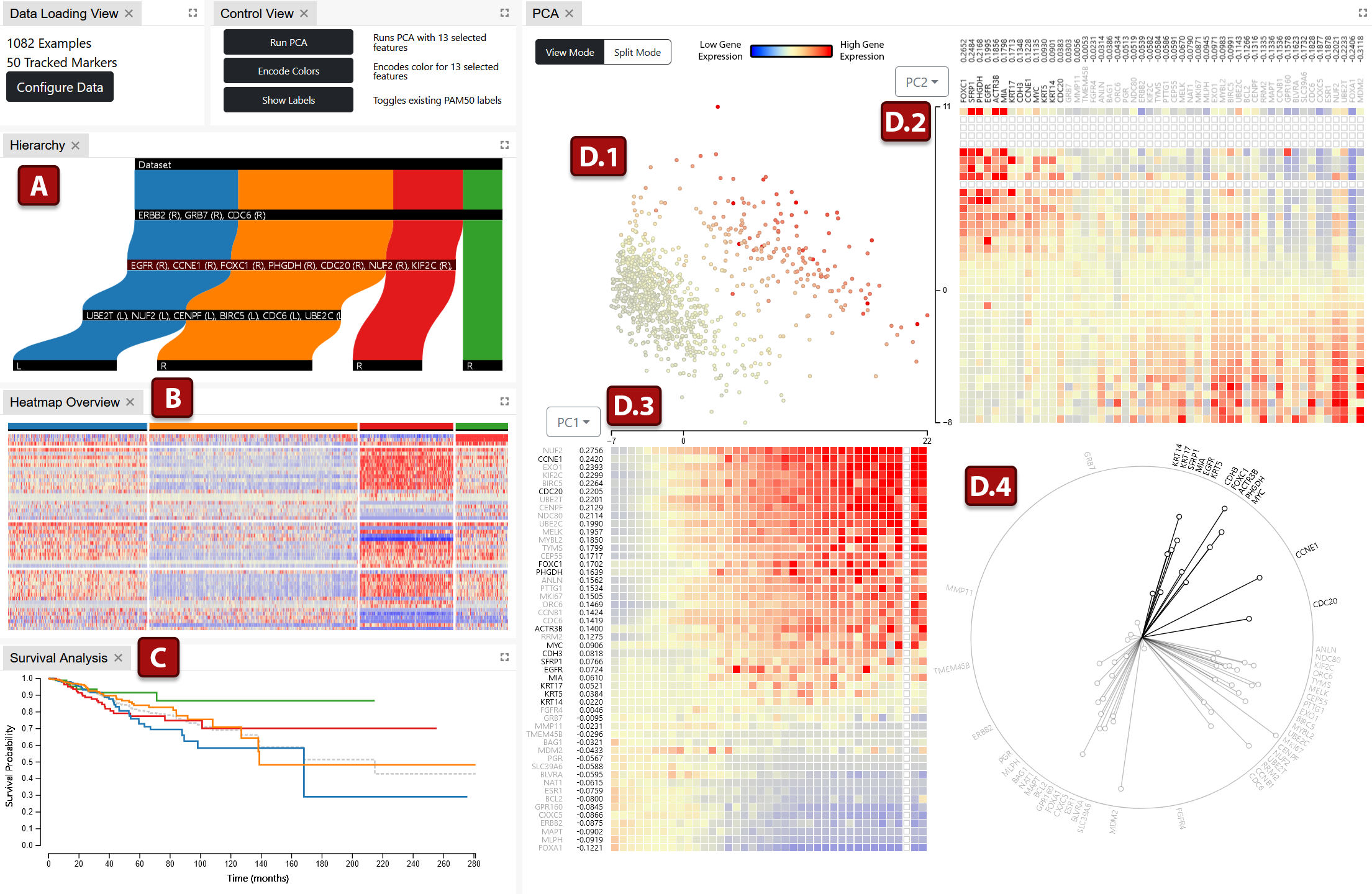}
  \vspace{-4 mm}
  \caption{
  An overview of the Vis-SPLIT tool.
  (A) The Hierarchical Overview is an abstract view of the current clusters.
  (B) The Heatmap Overview displays the patterns for the expression values of genes in each cluster.
  (C) The Survival Analysis View visualizes survival curves for each cluster. 
  (D) The PCA View allows users to view or split the selected node in the Hierarchical Overview, and includes
  (D.1) the Projection depicting individuals in 2D, placed based on genetic expression, 
  (D.2-D.3) axis-aligned heatmaps displaying the expression values of genes, 
  and (D.4) the Feature Loadings Plot showing current gene contributions.}
  \label{fig:teaser}
}

\abstract{
We propose an interactive visual analytics tool, Vis-SPLIT, for partitioning a population of individuals into groups with similar gene signatures.
Vis-SPLIT allows users to interactively explore a dataset and exploit visual separations to build a classification model for specific cancers.
The visualization components reveal gene expression and correlation to assist specific partitioning decisions, while also providing overviews for the decision model and clustered genetic signatures.
We demonstrate the effectiveness of our framework through a case study and evaluate its usability with domain experts.
Our results show that Vis-SPLIT can classify patients based on their genetic signatures to effectively gain insights into RNA sequencing data, as compared to an existing classification system.
} % end of abstract

\CCScatlist{
  \CCScatTwelve{Human-centered computing}{Visu\-al\-iza\-tion}{Visu\-al\-iza\-tion application domains}{Visual analytics}{}
}

\begin{document}


% \firstsection{Introduction}

% \maketitle

\firstsection{Introduction}
\maketitle

RNA-Sequencing (RNA-Seq) generates data about the abundance/expression of RNA molecules. This technique allows the identification of expression patterns that represent different cell states, which can have special diagnostic or prognostic value in cancer research. However, the success of this approach depends on the specificity of the context under consideration, such as normal tissue biology, immunogenic mutational burden, genetic etiology, or specific treatments. The earlier RNA studies required laborious manual assessment of the importance of individual genes in such context, aided by traditional readily-available hierarchical clustering techniques.

% Problems/Challenges
To cluster or analyze high-dimensional gene data, dimensionality reduction techniques have been commonly used. There are two types of dimensionality reduction techniques: linear methods such as Principal Component Analysis~(PCA)~\cite{pearson1901} and non-linear methods such as Multi-Dimensional Scaling~(MDS)~\cite{kruskal1964multidimensional} and t-Stochastic Neighbor Embedding~(t-SNE)~\cite{van2008visualizing}. While non-linear methods have been used successfully for capturing local distances between data points \cite{amir2013visne, hollt2016cytosplore}, they may require tools to explain them and rate their trustworthiness to generate better results~\cite{chatzimparmpas2020tvisne}.

Typical instances of RNA-Seq clustering tend to be static, so even relatively minor improvements such as adding or removing a feature in a cluster are prohibitively costly for subsequent reviewers or analysts, significantly limiting the pace of potential improvements via unplanned contributions. For example, the breast tumor RNA expression clusters were discovered using one-shot (model-free) hierarchical clustering and disseminated in static dendrograms and heatmaps~\cite{perou2000molecular,sorlie2001gene}. While the clusters have proven to carry a high degree of diagnostic and prognostic value in the ensuing years, progress on refinement of these results was slow. It was 8 years until an actual classifier derived from this discovery was developed, the PAM50 classifier~\cite{parker2009supervised}, and the corresponding clusters it classifies are essentially unimproved after its proposal.
Some machine learning techniques have found some success in improving classifications by training data-driven models~\cite{chen2018gsae,ding2018interpretable,lin2017using}.
However, these unsupervised, or ``black box'', techniques often lack transparency in their decisions and do not allow the incorporation of domain knowledge~\cite{kiselev2019challenges}.

To address these issues, we propose Vis-SPLIT (Visually Separable Plots formed from Linear, Iterated Technique), an interactive clustering framework that utilizes PCA to present users with easily explainable projections.
For our approach, we choose to use PCA due to its computational efficiency, interpretability, and the ability to visualize and interact with its inner workings \cite{shlens2014tutorial}, including the eigenvectors and eigenvalues of each Principal Component (PC).
In the proposed method, an analyst applies PCA to a working dataset iteratively to identify increasingly subtle clusters~\cite{sacha2017dimensionality}. Obvious partitions are made first, reducing the sizes of working clusters and ultimately revealing underlying patterns.

Vis-SPLIT provides four linked views that are designed to build a classification scheme for RNA-Seq data. Due to its iterative nature, only one projection is given at a time, along with other coordinated views to manipulate the projection into visually separable clusters.
The framework enables domain experts to incorporate domain knowledge by exploring and selecting high-dimensional features/genes. 
We demonstrate the usability of Vis-SPLIT through a case study on an RNA-Seq dataset of breast cancer patients and evaluate the effectiveness of Vis-SPLIT through domain experts’ feedback.

\section{Related Work}

Several clustering methods, including hierarchical methods~\cite{eisen1998cluster} and the $k$-means algorithm~\cite{tavazoie1999systematic} have been used to classify gene expression, but require tuning parameters and using appropriate methods for measuring similarity.

To address this issue, different interactive clustering methods have been successful when working with genetic data \cite{van2010interactive, seo2002interactively, mukhopadhyay2012interactive}. 
Van Long et al.~\cite{van2010interactive} presented an interactive, density-based hierarchical clustering method, which can deal with noises in microarray experiments. 
Mukhopadhyay et al.~\cite{mukhopadhyay2012interactive} proposed an interactive multiobjective clustering (IMOC) algorithm, learning from user decisions to refine its results.
Seo and Shneiderman~\cite{seo2002interactively} developed the Hierarchical Clustering Explorer for interactively exploring and visualizing large microarray data based on a hierarchical clustering method.
However, these methods often lack explanation of the clustering process. 

Some interactive tools utilize non-linear techniques to provide powerful data exploration features~\cite{somarakis2019imacyte, sabando2020chemva, hollt2017cyteguide}.
Somarakis et al.~\cite{somarakis2019imacyte} used t-SNE for their two-dimensional embeddings while H{\"o}llt et al.~\cite{hollt2017cyteguide} used a hierarchical variant, HSNE~\cite{pezzotti2016hierarchical}, to organize clusters into a radial hierarchy for exploration.
Sabando et al.~\cite{sabando2020chemva} trained a parametric feed-forward neural network to recreate the effects of t-SNE to be more fitting for their purpose.
While t-SNE based methods easily identify neighborhoods and provide separable projections, they do not explain projections in terms of the input features.

For interpretability and efficient clustering, many methods utilize linear dimensional reduction methods such as PCA when working with high-dimensional data.
PCA has been used successfully in the analysis of genetic data~\cite{treutlein2014reconstructing, usoskin2015unbiased, hu2019panoview}, but often lacks interactive configuration or is implemented and configured manually, targeted to a specific dataset.
This limits the flexibility of an approach to be used with different datasets.

There are more generally applicable PCA systems~\cite{jeong2009ipca, garrison2021dimlift} that provide additional insight into a dataset by exposing certain features of the algorithm.
iPCA~\cite{jeong2009ipca} allows the user to modify dimensional contribution and visualize the eigenvectors within the PCs.
DimLift~\cite{garrison2021dimlift} utilizes an iterative approach based on Factor Analysis of Mixed Data (FAMD), identifying obvious feature correlations first so that hidden patterns can be uncovered.
While useful tools, these methods focus on data and algorithm exploration or hypothesis formation, while Vis-SPLIT allows users to quickly build classification models for gene expression data.
\section{Design Requirements} 

RNA-Seq is a biological tissue measurement procedure and raw-data processing technique that ascertains the abundance (or ``expression'') of RNA transcripts in the sample for each gene. RNA-Seq is now routinely performed on tumor and adjacent tissue samples from cancer patients. The technique of honing in on expression patterns -- representing cell states -- with special diagnostic or prognostic value in research cohorts has been very successful, provided that the context under consideration is specific enough, e.g. with respect to tumor origin site, metastatic status, immunogenic mutational burden, specific genetic etiology, or specific treatments. However, discovering/classifying expression patterns is challenging due to the presence of high dimensionality~\cite{Vasighizaker:2022} and noises~\cite{Ficklin:2017}.
Based on discussions with two biomedical researchers, we identified several design requirements (DRs) of Vis-SPLIT for classifying types of cancer based on genetic expression.

\textbf{DR1}. \textit{Provide an overview of the distribution and features of individuals based on clusters:} Domain experts are interested in the characteristics of each cluster so that they can identify significant features/genes of each cluster/group, explore how common/rare each cluster is, and understand an overall structure of the user-defined classification.

\textbf{DR2}. \textit{Identify patterns of activation or inactivation for different genes or gene groups:}
In order to define meaningful clusters, an analyst needs to be able to identify genetic patterns that distinguish some individuals from the others. These patterns must be apparent within the working set of individuals so that the analyst can define classification rules to exploit the pattern's presence.

\textbf{DR3}. \textit{Compare the similarities of gene group activation across clusters:} 
It is common to analyze genetic distinction in terms of gene groups, or sets of genes whose collective activation is significant for the purposes of classifying and understanding an individual. These gene groups must be apparent and comparable across all clusters to confirm the results of a classification.

\textbf{DR4}. \textit{Understand the link between clusters and diseases:} 
Specific sub-types or classes of disease can be associated with each cluster, often defined by its highly activated gene groups and/or prior domain knowledge. Domain experts are interested in the survival rates and prognosis of these diseases.
\section{Framework}
As illustrated in \hyperref[fig:teaser]{Fig.~\ref{fig:teaser}}, Vis-SPLIT has four main views: (A) the Hierarchical Overview, (B) the Heatmap Overview, (C) the Survival Analysis View, and (D) the PCA View. 

\subsection{Hierarchical Overview}\label{hierarchical-overview}
The Hierarchical Overview (\hyperref[fig:teaser]{Fig.~\ref{fig:teaser}A}) shows the entire dataset being partitioned into individual clusters. In the Hierachical Overview, we can discover the distribution of patients with similar genetic signatures across clusters (\textbf{DR1}). This view serves as a visual representation of the classification model being built.

Visually, the plot resembles a top-down Sankey Diagram, with each rectangular node representing a separation in the data defined by user interactions with the PCA view.
In other words, the top of the diagram represents the whole dataset, with movement down the tree corresponding to iterative partitions in the data, eventually resulting in final clusters in the leaf nodes.

The width of nodes and bands corresponds to the number of individuals present in that portion of the classification model.
Colors are assigned to each cluster as partitions are made, and can be traced down to the leaf node (representing a final cluster) or can be used to see the relative size difference in any parenting super-clusters or the dataset as a whole.
A list of features (genes) can be found on each partitioning node, representing the features that were determined to be most different between the resulting clusters of a partition.
A given feature $i$ will be shown here if the following threshold is met:
\[
|\mu_i^{a} - \mu_i^{b}| \geq \sigma_{avg}
\]
where $\mu_i^{a}$ and $\mu_i^{b}$ represent the mean values of a feature $i$ among the two resulting clusters, $a$ and $b$, and $\sigma_{avg}$ represents $1$ standard deviation from the mean of all differences in feature values:
\[
\sigma_{avg} = \sqrt{\frac{\sum(d_i - \mu_{d})^2}{N}}
\]
where $d_i$ is the difference between the average value of feature $i$ across cluster $a$ and the average value of feature $i$ across cluster $b$, $\mu_d$ is the average of differences $d_i$ for all $i$ features, and $N$ is the number of features.

\subsection{Heatmap Overview}\label{heatmap-overview}
The Heatmap Overview (\hyperref[fig:teaser]{Fig.~\ref{fig:teaser}B}) shows an overview of all individuals and their genetic signatures. This allows comparison of activation in different gene groups across all clusters (\textbf{DR3})

Every individual is shown as a column, and every feature is represented as a row. The color of each bar encodes a gene expression value, blue (negative values) to yellow (zero) to red (positive values).
As partitions are made in the data, this heatmap will reorganize individuals into separable vertical bands, corresponding to the cluster bands of the Hierarchical Overview.
The corresponding cluster color is displayed at the top of each cluster in the Heatmap Overview. 
Additionally, any features that are identified as ``important'' for a given partition in the Hierarchical Overview will be grouped into a horizontal band, ordered internally by the highest expression.

\subsection{Survival Analysis View}\label{desc:survival-curves}
The Survival Analysis view (\hyperref[fig:teaser]{Fig.~\ref{fig:teaser}C}) depicts the relationship between formed clusters and diseases (\textbf{DR4}).
This view provides a summary of probabilities that individuals for each cluster will survive up to a specific time.
Kaplan-Meier analysis~\cite{kaplan1958nonparametric} is applied to the current clusters, and a curve is shown for each, colored to match that cluster's band in the Hierarchical Overview.
A baseline curve is also shown as a dotted gray line, representing the dataset as a whole.

\subsection{PCA View}
The PCA View (\hyperref[fig:teaser]{Fig.~\ref{fig:teaser}D}) is designed to display activation patterns for different groups of genes (\textbf{DR2}). In the PCA View, an analyst can identify feature trends and make meaningful partitions in the data. We use the same color encoding as the Heatmap Overview.

\subsubsection{Projection}
The Projection (\hyperref[fig:teaser]{Fig.~\ref{fig:teaser}-D.1}), shows the current PCA projection.
The projection will only include individuals from the selected node in the Hierarchical Overview and will utilize all features (genes) by default, though a limited feature space will be used if any features have been specifically selected. 
An analyst can choose any two Principle Components (PCs) for the Projection's axes, and can explore the configurations that utilize the most distinguishing features.
Each individual is represented as a point and is placed relative to the selected PCs.
In the Projection, an analyst can draw a divider line to partition the data based on an observed visual separation.
Color can also be encoded on points based on selected features.
This is done by averaging selected feature values for each individual and coloring them using the same blue-yellow-red color scale used in all of Vis-SPLIT's heatmaps. 
Additionally, if there is any existing classification for the dataset, it can be viewed for comparison by toggling an overlay showing categorical colors and a legend.  

\subsubsection{Heatmaps}\label{heatmaps}
For both selected PCs in the Projection, a heatmap is aligned, seen in \hyperref[fig:teaser]{Fig.~\ref{fig:teaser}-D.2} and \hyperref[fig:teaser]{Fig.~\ref{fig:teaser}-D.3}.
Each heatmap is divided into bins that encapsulate the aligned points in the Projection. 
For the bottom heatmap, each row indicates a feature/gene and each column represents a bin containing all individuals stretching upwards into the scatterplot. Features are sorted based on values within the PC's eigenvector, prioritizing the most contributing genes for a given projection. For the right heatmap, the rows and columns are swapped.

Each cell is colored based on the average value of the given feature for the contained individuals within the given heatmap bin. 
Heatmap bins that do not hold any individuals will instead show gray outlined boxes for each feature.

Feature names are listed to the sides of the heatmap along with their respective values in the heatmap's PC. 
Features can be selected by clicking these labels, which updates the global feature selection across all the visual components in the PCA View.
In the case that the user has selected any features, the selected features' labels remain black while any others are displayed in gray.

\subsubsection{Feature Loadings}\label{feature-loadings}
The Feature Loadings plot (\hyperref[fig:teaser]{Fig.~\ref{fig:teaser}-D.4}) shows the influence of each feature along the selected PCs in the Projection~\cite{holland2008principal}. 
Each vector represents a feature with its length and direction indicating its influence in the Projection.
A similar vector direction of features can indicate a correlation between them.
Features' labels are placed along an outside circle to reduce visual clutter. Additionally, spacing forces are applied to the labels to reduce overlap. 
An analyst can select features by clicking their circle or label or by brushing.
Unselected features and their vectors are grayed out and, like with the aligned heatmaps, any selection of features is global to the PCA View.

\section{Case Study}\label{use-case}

We demonstrate a classification case of our workflow to group patients with breast cancer. In this case, we used the PanCancer Atlas breast cancer dataset~\cite{tcgapancancer}.
This dataset tracks 50 genetic markers across 1082 individuals, which have been previously classified through the PAM50 test~\cite{Wallden:2015} as BRCA\_LumA, BRCA\_LumB, BRCA\_Basal, BRCA\_Her2, BRCA\_Normal, or none.

\begin{figure}
    \centering
    \includegraphics[width=\linewidth]{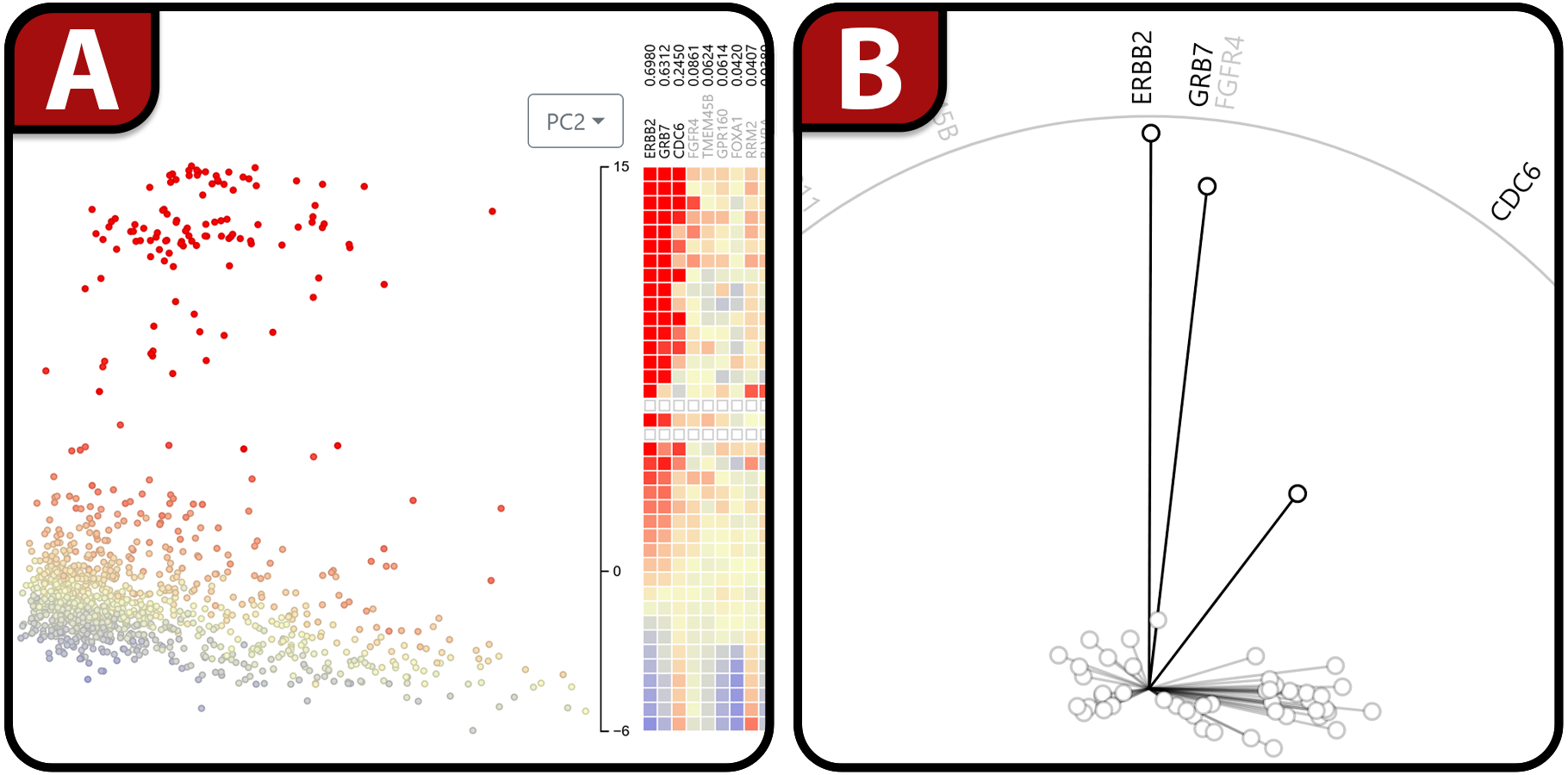}
    \vspace{-7 mm}
    \caption{An example of (A) identifying a gene partition based on (B) selected features in the Feature Loadings plot. 
    }
    \vspace{-4 mm}
    \label{fig:first-plot}
\end{figure}

To classify the dataset better than the PAM50, first, an analyst wants to divide the entire dataset into two parts through the Projection (\hyperref[fig:first-plot]{Fig.~\ref{fig:first-plot}A}). To examine the features/genes of each group in detail, the analyst explores a heatmap for the second PC, where a few genes seem to be much more positive towards the top of the scatterplot: ERBB2, GRB7 and CDC6~(\textbf{DR2}). Additionally, the Feature Loadings plot (\hyperref[fig:first-plot]{Fig.~\ref{fig:first-plot}B}) shows these features have a much stronger influence than the other features.

The analyst selects these seemingly correlated features from the Feature Loadings plot and encodes color to the Projection, confirming that individuals in the top part of the Projection have high expressions for these genes (\hyperref[fig:first-plot]{Fig.~\ref{fig:first-plot}A}).
Using their domain knowledge, the analyst knows these features to correlate to HER2 positive breast cancers, having higher levels of HER2 protein.
Based on this finding, the analyst divides the dataset into two clusters. 

The width of vertical bands in the Hierarchical and Heatmap Overviews reveal that about $10\%$ of individuals fall within the new HER2 group~(\textbf{DR1}).
The analyst can also see that the three significant genes they identified earlier have been separated and raised to the top of the Heatmap Overview, highlighting the major difference between these two clusters' genetic signatures~(\textbf{DR3}).
The analyst also notes that the HER2 group has a flatter survival curve in the Survival Analysis View, from which hypotheses may be formed about the prognosis of disease found within these individuals (\textbf{DR4}).

Next, to further classify individuals who don't belong to the HER2 group, the analyst chooses the larger leaf node in the Hierarchy Overview.
Again a separation is noticeable in the Projection, with one cluster near the bottom-left and another near the top-right~(\hyperref[fig:teaser]{Fig.~\ref{fig:teaser}D}).
The analyst selects all features pointing towards these directions, and runs PCA again.
The new projection is made using only about half of the total features, making it easier to focus on and linearly divide the clusters previously observed.

\begin{figure}
    \centering
    \includegraphics[width=\linewidth]{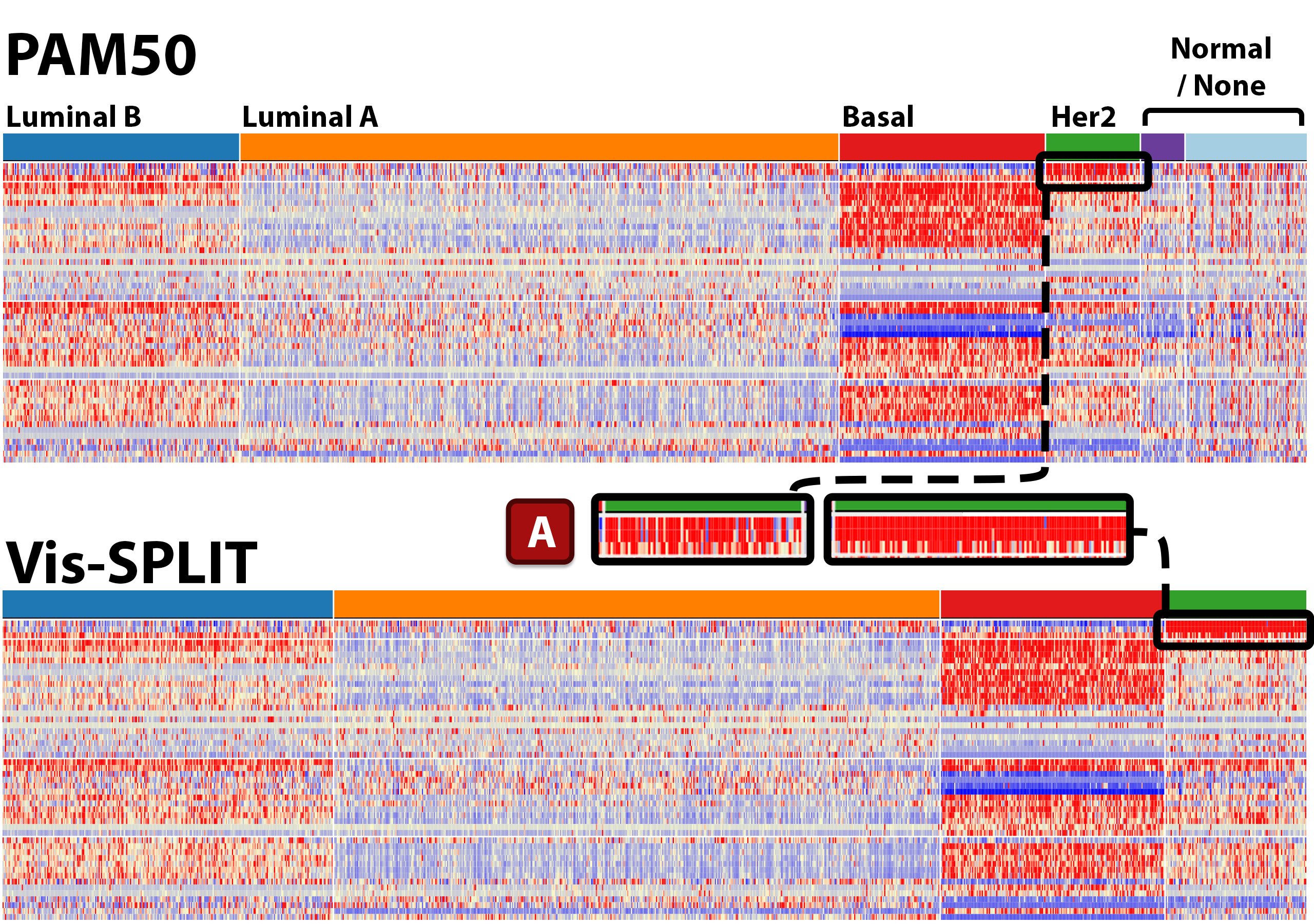}
    \vspace{-7 mm}
    \caption{A comparison of classification from the resulting Heatmap Overview of the PAM50 (top) and Vis-SPLIT (bottom), and (A) some regions where the difference is visible.
    }
    \vspace{-4 mm}
    \label{fig:comparison}
\end{figure}

The analyst continues to look for feature patterns and splits the data until they can no longer make any meaningful partitions. 
Once the analyst has completed their interactions, they review the resulting model in the Overviews and Survival Analysis View.
The produced Vis-SPLIT classification can then be compared with that of the PAM50 test (\hyperref[fig:comparison]{Fig.~\ref{fig:comparison}}).
While both heatmaps reveal cohesive genetic signatures across four distinct clusters, Vis-SPLIT classifies all examples, whereas the PAM50 test leaves some individuals within the less descriptive classifications of ``BRCA\_Normal'' and ``none''.
Though some of these previously unclassified individuals may be important to recognize as outliers, many have genetic signatures which strongly align with a formed cluster.
Additionally, some genetic patterns are more clearly represented within Vis-SPLIT clusters.
One example of these patterns is the top feature group within the HER2 positive clusters, which contains more overwhelmingly positive values for the three identified genes~(\hyperref[fig:comparison]{Fig.~\ref{fig:comparison}A}).

\subsection{Expert Feedback}

Our Vis-SPLIT was reviewed by two biomedical researchers to evaluate the usefulness of the proposed framework. 
After domain experts were given time to use Vis-SPLIT, along with some guidance on its use, they provided overall positive feedback, claiming that the tool was intuitive and produced an easily interprettable model.
They appreciated how fast it was to discover clusters and build a model, noting on first look that it ``truly felt most sensible to split first [based on the HER2 group], because it was the most salient [within the Projection].''
They liked using the Projection to extract clusters such as this one and one resembling PAM50's Basal group, and enjoyed the explanations provided by the aligned heatmaps and Loadings plot. 
For less significant distinctions, such as that between the orange and blue clusters from \hyperref[fig:comparison]{Fig.~\ref{fig:comparison}}, the domain experts favored patterns seen in the aligned heatmaps and Loadings plot, which revealed a group of genes that spanned continuously from low expression to high expression across the cluster.

The experts also liked the ability to overlay existing classifications directly on their projections. 
When viewing PAM50 classes through this option, one expert noticed that some individuals seemed misclassified as HER2 by the PAM50 system.
They reached this conclusion by encoding the genes known to correlate with this cancer onto the Projection, observing low representation on specific individuals and then overlaying the PAM50 classification.
They felt it was useful to be able to identify these inconsistencies in the genetic expressions within PAM50 classes and were more confident that the classifications made through Vis-SPLIT better grouped similar genetic signatures. 

The experts did note that the system did not provide any statistical analysis allowing more in-depth comparison of survival curves.
Additionally they mentioned a desire for more robust navigation in the Hierarchical Overview to return back to the previous states of the application, essentially being able to ``undo'' and/or review partitioning decisions.

% --Full Paper--
% \section{Discussion and Future Work}
% TODO
% \subsection{Discussion}
% \subsection{Limitations and future work}

\section{Conclusion}
We presented Vis-SPLIT, a novel framework for interactively clustering RNA-Seq datasets.
We provide several techniques to cluster similar genes and identify groups of individuals linked to the sub-types of a disease.
Finally, a case study and accompanying expert feedback is presented to demonstrate the tool's use.

There are several limitations of our current framework, which will be addressed in the future. First, although Vis-SPLIT has analyzed and visualized up to $2,000$ individuals, any more individuals results in the Heatmap Overview becoming increasingly crowded and losing the ability to identify individuals, especially once there is less than a pixel dedicated to each.
The addition of other summarization methods~\cite{Albers:2011} and a zoom feature can alleviate this issue.
With more features another limitation is seen in the Loadings Plot, where feature labels may stray further from the direction their corresponding vector.
Finally, the approach could be strengthened with measurements of certainty for an example's classification within the model.

Vis-SPLIT can assist cancer researchers in the exploration of data to build new models and compare them with existing baselines.
Outputted models can be used directly or can be summarized into simpler decision trees based on significant gene expressions.

% The authors wish to thank A, B, and C. This work was supported in part by
% a grant from XYZ.}
% This work was supported in part by MSK Cancer Center Support Grant/Core Grant (P30 CA008748).
% }

\bibliographystyle{abbrv}
% \bibliographystyle{abbrv-doi}
% \bibliographystyle{abbrv-doi-narrow}
% \bibliographystyle{abbrv-doi-hyperref}
%\bibliographystyle{abbrv-doi-hyperref-narrow}

% Uncomment to test if bibliography is under 1 page
% \clearpage

\bibliography{template}

\begin{thebibliography}{10}

\bibitem{Albers:2011}
D.~Albers, C.~Dewey, and M.~Gleicher.
\newblock Sequence surveyor: Leveraging overview for scalable genomic alignment
  visualization.
\newblock {\em IEEE Transactions on Visualization and Computer Graphics},
  17(12):2392--2401, 2011.

\bibitem{amir2013visne}
E.-a.~D. Amir, K.~L. Davis, M.~D. Tadmor, E.~F. Simonds, J.~H. Levine, S.~C.
  Bendall, D.~K. Shenfeld, S.~Krishnaswamy, G.~P. Nolan, and D.~Pe'er.
\newblock visne enables visualization of high dimensional single-cell data and
  reveals phenotypic heterogeneity of leukemia.
\newblock {\em Nature biotechnology}, 31(6):545--552, 2013.

\bibitem{chatzimparmpas2020tvisne}
A.~Chatzimparmpas, R.~M. Martins, and A.~Kerren.
\newblock {t-viSNE}: Interactive assessment and interpretation of t-sne
  projections.
\newblock {\em IEEE Transactions on Visualization and Computer Graphics},
  26(8):2696--2714, 2020.

\bibitem{chen2018gsae}
H.-I.~H. Chen, Y.-C. Chiu, T.~Zhang, S.~Zhang, Y.~Huang, and Y.~Chen.
\newblock Gsae: an autoencoder with embedded gene-set nodes for genomics
  functional characterization.
\newblock {\em BMC systems biology}, 12(8):45--57, 2018.

\bibitem{ding2018interpretable}
J.~Ding, A.~Condon, and S.~P. Shah.
\newblock Interpretable dimensionality reduction of single cell transcriptome
  data with deep generative models.
\newblock {\em Nature communications}, 9(1):2002, 2018.

\bibitem{eisen1998cluster}
M.~B. Eisen, P.~T. Spellman, P.~O. Brown, and D.~Botstein.
\newblock Cluster analysis and display of genome-wide expression patterns.
\newblock {\em Proceedings of the National Academy of Sciences},
  95(25):14863--14868, 1998.

\bibitem{Ficklin:2017}
S.~P. Ficklin, L.~J. Dunwoodie, W.~L. Poehlman, C.~Watson, K.~E. Roche, and
  F.~A. Feltus.
\newblock Discovering condition-specific gene co-expression patterns using
  gaussian mixture models: A cancer case study.
\newblock {\em Scientific Reports}, 7(1):8617, 2017.

\bibitem{garrison2021dimlift}
L.~Garrison, J.~M{\"u}ller, S.~Schreiber, S.~Oeltze-Jafra, H.~Hauser, and
  S.~Bruckner.
\newblock Dimlift: Interactive hierarchical data exploration through
  dimensional bundling.
\newblock {\em IEEE Transactions on Visualization and Computer Graphics},
  27(6):2908--2922, 2021.

\bibitem{holland2008principal}
S.~M. Holland.
\newblock Principal components analysis (pca).
\newblock {\em Department of Geology, University of Georgia, Athens, GA}, pages
  30602--2501, 2008.

\bibitem{hollt2016cytosplore}
T.~H{\"o}llt, N.~Pezzotti, V.~van Unen, F.~Koning, E.~Eisemann, B.~Lelieveldt,
  and A.~Vilanova.
\newblock Cytosplore: interactive immune cell phenotyping for large single-cell
  datasets.
\newblock 35 issue 3:171--180, 2016.

\bibitem{hollt2017cyteguide}
T.~H{\"o}llt, N.~Pezzotti, V.~van Unen, F.~Koning, B.~P. Lelieveldt, and
  A.~Vilanova.
\newblock Cyteguide: Visual guidance for hierarchical single-cell analysis.
\newblock {\em IEEE Transactions on Visualization and Computer Graphics},
  24(1):739--748, 2017.

\bibitem{hu2019panoview}
M.-W. Hu, D.~W. Kim, S.~Liu, D.~J. Zack, S.~Blackshaw, and J.~Qian.
\newblock Panoview: An iterative clustering method for single-cell rna
  sequencing data.
\newblock {\em PLoS computational biology}, 15(8):e1007040, 2019.

\bibitem{jeong2009ipca}
D.~H. Jeong, C.~Ziemkiewicz, B.~Fisher, W.~Ribarsky, and R.~Chang.
\newblock {iPCA}: An interactive system for pca-based visual analytics.
\newblock {\em Computer Graphics Forum}, 28(3):767--774, 2009.

\bibitem{kaplan1958nonparametric}
E.~L. Kaplan and P.~Meier.
\newblock Nonparametric estimation from incomplete observations.
\newblock {\em Journal of the American statistical association},
  53(282):457--481, 1958.

\bibitem{kiselev2019challenges}
V.~Y. Kiselev, T.~S. Andrews, and M.~Hemberg.
\newblock Challenges in unsupervised clustering of single-cell rna-seq data.
\newblock {\em Nature Reviews Genetics}, 20(5):273--282, 2019.

\bibitem{kruskal1964multidimensional}
J.~B. Kruskal.
\newblock Multidimensional scaling by optimizing goodness of fit to a nonmetric
  hypothesis.
\newblock {\em Psychometrika}, 29(1):1--27, 1964.

\bibitem{lin2017using}
C.~Lin, S.~Jain, H.~Kim, and Z.~Bar-Joseph.
\newblock Using neural networks for reducing the dimensions of single-cell
  rna-seq data.
\newblock {\em Nucleic acids research}, 45(17):e156--e156, 2017.

\bibitem{mukhopadhyay2012interactive}
A.~Mukhopadhyay, U.~Maulik, and S.~Bandyopadhyay.
\newblock An interactive approach to multiobjective clustering of gene
  expression patterns.
\newblock {\em IEEE Transactions on Biomedical Engineering}, 60(1):35--41,
  2012.

\bibitem{parker2009supervised}
J.~S. Parker, M.~Mullins, M.~C. Cheang, S.~Leung, D.~Voduc, T.~Vickery,
  S.~Davies, C.~Fauron, X.~He, Z.~Hu, et~al.
\newblock Supervised risk predictor of breast cancer based on intrinsic
  subtypes.
\newblock {\em Journal of clinical oncology}, 27(8):1160, 2009.

\bibitem{pearson1901}
K.~Pearson.
\newblock Liii. on lines and planes of closest fit to systems of points in
  space.
\newblock {\em The London, Edinburgh, and Dublin philosophical magazine and
  journal of science}, 2(11):559--572, 1901.

\bibitem{perou2000molecular}
C.~M. Perou, T.~S{\o}rlie, M.~B. Eisen, M.~Van De~Rijn, S.~S. Jeffrey, C.~A.
  Rees, J.~R. Pollack, D.~T. Ross, H.~Johnsen, L.~A. Akslen, et~al.
\newblock Molecular portraits of human breast tumours.
\newblock {\em nature}, 406(6797):747--752, 2000.

\bibitem{pezzotti2016hierarchical}
N.~Pezzotti, T.~H{\"o}llt, B.~Lelieveldt, E.~Eisemann, and A.~Vilanova.
\newblock Hierarchical stochastic neighbor embedding.
\newblock {\em Computer Graphics Forum}, 35(3):21--30, 2016.

\bibitem{sabando2020chemva}
M.~V. Sabando, P.~Ulbrich, M.~Selzer, J.~By{\v{s}}ka, J.~Mi{\v{c}}an,
  I.~Ponzoni, A.~J. Soto, M.~L. Ganuza, and B.~Kozl{\'\i}kov{\'a}.
\newblock Chemva: interactive visual analysis of chemical compound similarity
  in virtual screening.
\newblock {\em IEEE Transactions on Visualization and Computer Graphics},
  27(2):891--901, 2020.

\bibitem{sacha2017dimensionality}
D.~Sacha, L.~Zhang, M.~Sedlmair, J.~A. Lee, J.~Peltonen, D.~Weiskopf, S.~C.
  North, and D.~A. Keim.
\newblock Visual interaction with dimensionality reduction: A structured
  literature analysis.
\newblock {\em IEEE Transactions on Visualization and Computer Graphics},
  23(1):241--250, 2017.

\bibitem{seo2002interactively}
J.~Seo and B.~Shneiderman.
\newblock Interactively exploring hierarchical clustering results [gene
  identification].
\newblock {\em Computer}, 35(7):80--86, 2002.

\bibitem{shlens2014tutorial}
J.~Shlens.
\newblock A tutorial on principal component analysis.
\newblock {\em arXiv preprint arXiv:1404.1100}, 2014.

\bibitem{somarakis2019imacyte}
A.~Somarakis, V.~Van~Unen, F.~Koning, B.~Lelieveldt, and T.~H{\"o}llt.
\newblock Imacyte: visual exploration of cellular micro-environments for
  imaging mass cytometry data.
\newblock {\em IEEE transactions on visualization and computer graphics},
  27(1):98--110, 2019.

\bibitem{sorlie2001gene}
T.~S{\o}rlie, C.~M. Perou, R.~Tibshirani, T.~Aas, S.~Geisler, H.~Johnsen,
  T.~Hastie, M.~B. Eisen, M.~Van De~Rijn, S.~S. Jeffrey, et~al.
\newblock Gene expression patterns of breast carcinomas distinguish tumor
  subclasses with clinical implications.
\newblock {\em Proceedings of the National Academy of Sciences},
  98(19):10869--10874, 2001.

\bibitem{tavazoie1999systematic}
S.~Tavazoie, J.~D. Hughes, M.~J. Campbell, R.~J. Cho, and G.~M. Church.
\newblock Systematic determination of genetic network architecture.
\newblock {\em Nature genetics}, 22(3):281--285, 1999.

\bibitem{tcgapancancer}
{The Cancer Genome Atlas (TCGA) Research Network}.
\newblock Pan-cancer atlas: Breast invasive carcinoma, 2018.
\newblock data retrieved through cBioPortal,
  \url{http://www.cbioportal.org/study/summary?id=brca_tcga_pan_can_atlas_2018}.

\bibitem{treutlein2014reconstructing}
B.~Treutlein, D.~G. Brownfield, A.~R. Wu, N.~F. Neff, G.~L. Mantalas, F.~H.
  Espinoza, T.~J. Desai, M.~A. Krasnow, and S.~R. Quake.
\newblock Reconstructing lineage hierarchies of the distal lung epithelium
  using single-cell rna-seq.
\newblock {\em Nature}, 509(7500):371--375, 2014.

\bibitem{usoskin2015unbiased}
D.~Usoskin, A.~Furlan, S.~Islam, H.~Abdo, P.~L{\"o}nnerberg, D.~Lou,
  J.~Hjerling-Leffler, J.~Haeggstr{\"o}m, O.~Kharchenko, P.~V. Kharchenko,
  et~al.
\newblock Unbiased classification of sensory neuron types by large-scale
  single-cell rna sequencing.
\newblock {\em Nature neuroscience}, 18(1):145--153, 2015.

\bibitem{van2008visualizing}
L.~Van~der Maaten and G.~Hinton.
\newblock Visualizing data using t-sne.
\newblock {\em Journal of machine learning research}, 9(11), 2008.

\bibitem{van2010interactive}
T.~Van~Long and L.~Linsen.
\newblock Interactive exploration of hierarchical density clusters in gene
  expression data.
\newblock {\em 2010 Second International Conference on Knowledge and Systems
  Engineering}, pages 20--27, 2010.

\bibitem{Vasighizaker:2022}
A.~Vasighizaker, S.~Danda, and L.~Rueda.
\newblock Discovering cell types using manifold learning and enhanced
  visualization of single-cell rna-seq data.
\newblock {\em Scientific Reports}, 12(1):120, 2022.

\bibitem{Wallden:2015}
B.~Wallden, J.~Storhoff, T.~Nielsen, N.~Dowidar, C.~Schaper, S.~Ferree, S.~Liu,
  S.~Leung, G.~Geiss, J.~Snider, T.~Vickery, S.~R. Davies, E.~R. Mardis,
  M.~Gnant, I.~Sestak, M.~J. Ellis, C.~M. Perou, P.~S. Bernard, and J.~S.
  Parker.
\newblock Development and verification of the pam50-based prosigna breast
  cancer gene signature assay.
\newblock {\em BMC Medical Genomics}, 8(1):54, 2015.

\end{thebibliography}
\end{document}